\documentclass[prl,print,showpacs,superscriptaddress,twocolumn]{revtex4}
\usepackage[dvips]{graphics,color}
\usepackage{amsfonts}
\usepackage{amsmath}
\usepackage{amssymb}
\usepackage{graphicx}

\begin{document}

\title{ Quantum resonant effect of the strongly-driven spin-boson model}
\author{Qifeng Liang}
\affiliation{State Key Laboratory of Quantum Optics and Quantum Optics Devices, College
of Physics and Electronic Engineering, Shanxi University, Taiyuan 030006, P.
R. China}
\affiliation{Department of Physics, Shaoxing University, Shaoxing 312000, P. R. China}
\author{Lixian Yu}
\affiliation{School of Physical Science and Technology, Suzhou University, Suzhou 215006,
P. R. China}
\affiliation{Department of Physics, Shaoxing University, Shaoxing 312000, P. R. China}
\author{Gang Chen \footnote{%
Corresponding author: chengang971@163.com}}
\affiliation{State Key Laboratory of Quantum Optics and Quantum Optics Devices, College
of Physics and Electronic Engineering, Shanxi University, Taiyuan 030006, P.
R. China}
\affiliation{Department of Physics, Shaoxing University, Shaoxing 312000, P. R. China}
\author{Suotang Jia}
\affiliation{State Key Laboratory of Quantum Optics and Quantum Optics Devices, College
of Physics and Electronic Engineering, Shanxi University, Taiyuan 030006, P.
R. China}

\begin{abstract}
In this paper we discuss both analytically and numerically the rich quantum
dynamics of the spin-boson model driven by a time-independent field of
photon. Interestingly, we predict a new Rabi oscillation, whose period is
inversely proportional to the driving amplitude. More surprisingly, some
nonzero resonant peaks are found for some special values of the \emph{strong}
driving regime. Moreover, for the different resonant positions, the peaks
have different values. Thus, an important application of this resonance
effect is to realize the precision measurement of the relative parameters in
experiment. We also illustrate that this resonant effect arises from the
interference of the nontrivial periodic phase factors induced by the
evolution of the coherent states in two different subspaces. Our predictions
may be, in principle, observed in the solid-state cavity quantum
electrodynamics with the ultrastrong coupling if the driving magnitude of
the photon field is sufficiently large.
\end{abstract}

\maketitle

Recent experiments about the solid-state cavity quantum electrodynamics
including the Josephosn junctions and the semiconducting dots have reported
the ultrastrong atom-photon interaction, whose magnitude is the same order
as that of the photon frequency \cite{Gunter,Anappara,Todorov,Fedorov}. In
particular, the ratio $0.12$ between the coupling strength and the microwave
photon frequency has been achieved successfully in the flux-based circuit
quantum electrodynamics \cite{Niemczyk}, and maybe approach unit and even go
beyond due to the current efforts \cite{Bourassa,Peropadre}. It is quite
different from the optical cavity quantum electrodynamics with the strong
coupling that in this so-called ultrastrong coupling regime, the well-known
rotating-wave approximation breaks down. As a consequence, the system's
dynamics is governed by the spin-boson model with the counter-rotating
terms, rather than the solvable Jaynes-Cummings model \cite{JC}.
Importantly, due to the existence of the counter-rotating terms, the
spin-boson model has fascinating quantum dynamics beyond that of the
Jaynes-Cummings model. The exploration of the exotic quantum effects in the
spin-boson model has been now of great interests \cite%
{Crisp,Lamata,Zhu,Gerritsma,Larson,Larson1,JH,Hirokawa,Casanova}, but still
has an open problem.

In this Letter we investigate both analytically and numerically the rich
quantum dynamics of the spin-boson model driven by a time-independent field
of photon. Interestingly, we predict a new Rabi oscillation, whose period is
inversely proportional to the driving amplitude. More surprisingly, some
nonzero resonant peaks are found for some special values of the \emph{strong}
driving regime. Moreover, for the different resonant positions, the peaks
have different values. Thus, an important application of this resonance
effect is to realize the precision measurement of the relative parameters in
experiment. We also illustrate that this resonant effect arises from the
interference of the nontrivial periodic phase factors induced by the
evolution of the coherent states in two different subspaces. Our predictions
may be observed in current experiment setups of solid-state cavity quantum
electrodynamics with the ultrastrong coupling if the driving magnitude of
the photon field is sufficiently large. However, our predicted quantum
resonant effect disappears when the spin-boson model is driven by a field of
atom, even if the driving magnitude is very strong.

It has been known that the atom-photon interaction is governed by the
spin-boson model \cite{SC}
\begin{equation}
H_{SB}=\omega a^{\dagger }a+\frac{1}{2}\varepsilon \sigma _{z}+g\sigma
_{x}(a^{\dagger }+a),  \label{SB}
\end{equation}%
where $a^{\dagger }(a)$ is the creation (annihilation) operator for photon
with frequency $\omega $, $\sigma _{z}$ and $\sigma _{x}$ are the Pauli spin
operators, $\varepsilon $ is the atomic resonant frequency, $g$ is the
atom-photon coupling strength. In the optical cavity with the strong
coupling $g\sim 10^{-5}\omega $ \cite{JM}, the counter-rotating terms $%
\sigma _{+}a^{\dagger }$ and $\sigma _{-}a$ are usually eliminated by means
of the rotating-wave approximation and the solvable Jaynes-Cummings model $%
H_{JC}=\omega a^{\dagger }a+\frac{1}{2}\varepsilon \sigma _{z}+g(\sigma
_{-}a^{\dagger }+\sigma _{+}a)$ can perfectly describe the system's dynamics
\cite{JC}. However, in the microwave cavity with the solid-state artificial
atom, the coupling strength has reached the ultrastrong regime $g\sim
0.1\omega $ \cite{Niemczyk}. Moreover, this ratio can be well controlled,
for example in the circuit cavity quantum electrodynamics, by the gate
capacitance, the gate voltage, the inductive coupling, and a transmission
line resonator \cite{RJ}. In such a large ratio, the rotating-wave
approximation breaks down. As a result, the counter-rotating terms $\sigma
_{+}a^{\dagger }$ and $\sigma _{-}a$ must be taken into account. Moreover,
these terms $\sigma _{+}a^{\dagger }$ and $\sigma _{-}a$ play a crucial role
in the quantum dynamics of Hamiltonian (\ref{SB}). As will be demonstrated,
a novel quantum resonant effect for Hamiltonian (\ref{SB}) driven strongly
by a time-independent photon field $H_{p}=\Omega _{p}(a^{\dagger }+a)$ with $%
\Omega _{p}$ being the driving magnitude is predicted by discussing the
experimentally-measurable quantum dynamics of $\left\langle \sigma
_{z}(t)\right\rangle $.
\begin{figure}[tp]
\includegraphics[width=6.2cm]{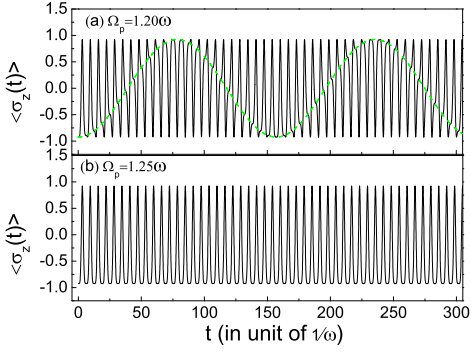}\newline
\caption{(Color online)The time-dependent quantum dynamics of $\left\langle
\protect\sigma _{z}(t)\right\rangle $ for the different driving magnitude $%
\Omega _{a}=1.2\protect\omega $ (a) and $\Omega _{a}=1.25\protect\omega $
(b) when $g=0.2\protect\omega $ and $\protect\varepsilon =0$. }
\label{fig1}
\end{figure}

With the time-independent driving field of photon, Hamiltonian (\ref{SB})
can be rewritten as $H_{P}=\omega a^{\dagger }a+\frac{1}{2}\varepsilon
\sigma _{z}+g\sigma _{x}(a^{\dagger }+a)+\Omega _{p}(a^{\dagger }+a)$. We
first consider the case of $\varepsilon =0$, in which the quantum dynamics
of $\left\langle \sigma _{z}(t)\right\rangle $ can be solved exactly for a
given initial state. In the following discussions, this initial state is
chosen as the eigenstate of Hamiltonian $H_{P}$ without the external driving
$(\Omega _{p}=0)$, namely, $H_{P}^{\Omega =0}=\omega a^{\dagger }a+g\sigma
_{x}(a^{\dagger }+a)$. Apparently, Hamiltonian $H_{P}^{\Omega =0}$ has the
property that $H_{P}^{\Omega =0}(-a^{\dagger },-a,-\sigma
_{x})=H_{P}^{\Omega =0}(a^{\dagger },a,\sigma _{x})$, which gives rise to
the degenerate ground states $\left\vert G_{\pm }(0)\right\rangle
=\left\vert z_{\pm }(0)\right\rangle \otimes \left\vert \pm \right\rangle $,
where $\left\vert z_{\pm }(0)\right\rangle =D(\mp \frac{g}{\omega }%
)\left\vert 0\right\rangle $ with $\left\vert 0\right\rangle $ being the
vacuum state of photon and $D(\xi )=\exp (\xi a^{\dagger }-\xi ^{\ast }a)$
being the displacement operator, and $\left\vert \pm \right\rangle $ are the
eigenstate of $\sigma _{x}$ \cite{Hwang}.

After the external driving field of photon is applied to control the
evolution of the coherent states, the degeneracy of Hamiltonian $%
H_{P}^{\Omega \neq 0}[=\omega a^{\dagger }a+g\sigma _{x}(a^{\dagger
}+a)+\Omega _{p}(a^{\dagger }+a)]$ breaks down. However, it still has a
conserved quantity $\sigma _{x}$, and correspondingly, its dynamics can be
well discussed in two subspaces with $\sigma _{x}=\pm 1$, whose effective
Hamiltonians are given respectively by $H_{\pm }=$ $\omega a^{\dagger
}a+g_{\pm }(a^{\dagger }+a)$ with $g_{\pm }=g\pm \Omega _{p}$. Under the
initial states $\left\vert G_{\pm }(0)\right\rangle $, the time-independent
wavefunctions for Hamiltonians $H_{\pm }$ can be obtained exactly by
\begin{equation}
\left\vert \varphi _{\pm }(t)\right\rangle =e^{i[\frac{g_{\pm }^{2}}{\omega }%
t-\frac{\Omega _{p}g_{\pm }}{\omega }\sin (\omega t)]}\left\vert z_{\pm
}(t)\right\rangle \otimes \left\vert \pm \right\rangle ,  \label{TW}
\end{equation}%
where the coherent states for any time is given by $\left\vert z_{\pm
}(t)\right\rangle =D(-g_{\pm }/\omega +\Omega _{p}e^{-i\omega t})\left\vert
0\right\rangle $. Based on these time-dependent wavefunctions $\left\vert
\varphi _{\pm }(t)\right\rangle $, the total wavefunction for Hamiltonian $%
H_{P}^{\Omega \neq 0}$ is given by $\left\vert \psi (t)\right\rangle
=\left\vert \varphi _{+}(t)\right\rangle -\left\vert \varphi
_{-}(t)\right\rangle $, which leads to the required dynamics $\left\langle
\sigma _{z}(t)\right\rangle =-\left\langle \varphi _{+}(t)\right\vert \sigma
_{z}\left\vert \varphi _{-}(t)\right\rangle -\left\langle \varphi
_{-}(t)\right\vert \sigma _{z}\left\vert \varphi _{+}(t)\right\rangle $ \cite%
{Note}, namely,
\begin{equation}
\left\langle \sigma _{z}(t)\right\rangle =-\exp (-\frac{2g^{2}}{\omega ^{2}}%
)\cos [\eta t-\xi \sin (\omega t)],  \label{FDT}
\end{equation}%
where $\eta =4g\Omega _{p}/\omega $ and $\xi =4g\Omega _{p}/\omega ^{2}$.

\begin{figure}[tp]
\includegraphics[width=6.0cm]{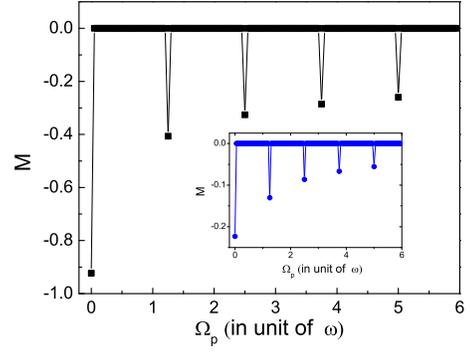}\newline
\caption{(Color online) The mean value $M$ of $\left\langle \protect\sigma %
_{z}(t)\right\rangle $ for a long time $T_{L}=50\protect\pi /\protect\omega $
as a function of the driving amplitude $\Omega _{p}$ when $g=0.2\protect%
\omega $ and $\protect\varepsilon =0$. Insert: The initial state is chosen
as the random state $\left\vert G(0)\right\rangle _{R}$.}
\label{fig2}
\end{figure}

Equation (\ref{FDT}), which is the main result of this Letter, describes the
quantum dynamics for \textit{any} driving magnitude. It can be found that a
coherent Rabi oscillation occurs here. Moreover, its period $T=\pi \omega
/(2g\Omega _{p})$, unlike the traditional Rabi oscillation of the
Jaynes-Cummings model, depends on the photon frequency $\omega $, the
coupling strength $g$. In particular, this period is inversely proportional
to the driving magnitude $\Omega _{p}$. If no driving is applied, this
coherent Rabi oscillation disappears. However, the photon number wave
packets can bounce back and forth along the same parity chains of
Hamiltonian (\ref{SB}), while producing collapse and revivals of the initial
population \cite{Casanova}. In general, the term $\xi \sin (\omega t)$ has
some effects on the time-dependent evolution of $\left\langle \sigma
_{z}(t)\right\rangle $. In Fig. 1(a), the dynamics of $\left\langle \sigma
_{z}(t)\right\rangle $ is plotted when $\Omega _{p}=1.2\omega $ and $%
g=0.2\omega $. In this figure, the dynamics depicted by the green line
results from the influence of the term $\xi \sin (\omega t)$. However, in a
special value for the \textit{strong} driving magnitude, the dynamics is
quite different. As shown in Fig. 1 (b), the green line disappears when $%
\Omega _{p}=1.25\omega $ with the same $g$. It seems that some exotic
quantum effects occur in this strong driving. In order to show this clearly,
we evaluate the mean value of $\left\langle \sigma _{z}(t)\right\rangle $
for a long time, namely, $M=\frac{1}{T_{L}}\int_{0}^{T_{L}}\left\langle
\sigma _{z}(t)\right\rangle dt$.

Interestingly, by means of the first-kind Bessel function $B_{m}(a)$, the
mean value $M$ is obtained by
\begin{equation}
M=-\exp (-\frac{2g^{2}}{\omega ^{2}})B_{m}(m)f(\Omega _{p}^{m}),  \label{MS}
\end{equation}%
where $f(\Omega _{p}^{m})$ is a jump function that satisfies $f(\Omega
_{p}^{m})=1$ for $\Omega _{p}^{m}=m\omega ^{2}/4g$ and $f(\Omega _{p}^{m})=0$
for $\Omega _{p}^{m}\neq m\omega ^{2}/4g$. Figure 2 is plotted numerically
the mean value $M$ for the time $T_{L}=50\pi /\omega $ as a function of the
driving amplitude $\Omega _{p}$. It is shown in this figure that for a
strong driving, namely, $\Omega _{p}^{m}=1.25m\omega $ ($g=0.2\omega $), a
nonzero mean value $M$ can be found, whereas $M=0$ for $\Omega _{p}^{m}\neq
1.25m\omega $. This phenomenon exhibits clearly that a novel quantum
resonant effect is predicted in the spin-boson model driven strongly by a
photon field. Figure 2 also shows that at the different resonant position $%
\Omega _{p}^{m}=1.25\omega $, $2.5\omega $,$\cdots $, the magnitudes of the
resonant peaks are different. Thus, it is very meaningful in experiment to
implement a precise determination about the relative parameters by measuring
$\Delta M_{m,m+i}=-\exp (-\frac{2g^{2}}{\omega ^{2}})[B_{m+i}(m+i)-B_{m}(m)]$
with $i=1,2,\cdots $. For example, in terms of the measuable $\Delta
M_{m,m+i}$, the couping strength can be obtained by
\begin{equation}
\frac{g}{\omega }=\sqrt{-\frac{1}{2}\ln \frac{\Delta M_{m,m+i}}{%
B_{m+i}(m+i)-B_{m}(m)}}.  \label{CS}
\end{equation}%
On the other hand, based on the different resonant positions, the driving
magnitude $\Omega _{p}$ can be also detected accurately. In the insert part
of Fig. 2, we also check that, if the initial state is chosen as a random
state $\left\vert G(0)\right\rangle
_{R}=\sum_{N=0}^{N=5}(C_{N}^{+}\left\vert N,+\right\rangle
+C_{N}^{-}\left\vert N,-\right\rangle )$, where $C_{N}^{\pm }=A_{\pm }\exp
[i2\pi B_{\pm }]$ with $A_{\pm }$ and $B_{\pm }$ being two random numbers
distributed in $[0,1)$ uniformly, the quantum resonant effect still remains
at the same positions.

\begin{figure}[tp]
\includegraphics[width=6.0cm]{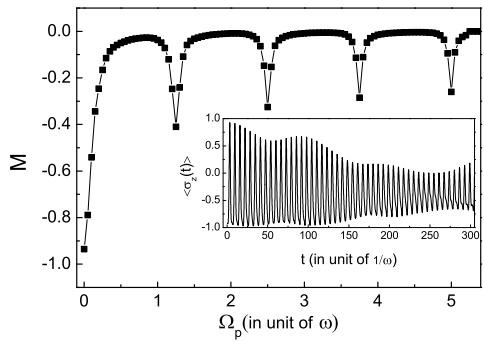}\newline
\caption{The mean value $M$ for a long time $T_{L}=50\protect\pi /\protect%
\omega $ as a function of the driving amplitude $\Omega _{p}$ when $g=0.2%
\protect\omega $ and $\protect\varepsilon =0.1\protect\omega $. Insert: The
quantum dynamics of $\left\langle \protect\sigma _{z}(t)\right\rangle $ for $%
g=0.2\protect\omega $, $\protect\varepsilon =0.1\protect\omega $ and $\Omega
=1.25\protect\omega $. }
\label{fig3}
\end{figure}

We now illustrate the physical explanation why this quantum resonant effect
can occur. In term of the eigenstates $\left\vert \pm \right\rangle $ of $%
\sigma _{x}$, Hamiltonian $H_{P}$ with $\varepsilon =0$ has two subspaces,
whose corresponding energies can be obtained exactly by $E_{\pm }=\omega
N_{\pm }-g_{\pm }^{2}/\omega $ with $N_{\pm }$ being the photon numbers for $%
H_{\pm }$. Therefore, the gap between the branches of energy is given by $%
\Delta E_{\pm }=E_{+}-E_{-}=\omega m-4\Omega _{p}g/\omega $. For a weak
driving, nothing happens in the gap $\Delta E_{\pm }$. However, for the
strong driving, the gap becomes zero at the resonant positions $\Omega
_{p}^{m}$. It means that the resonant effect perhaps has a correspondence on
the crossing between two branches of energy. An important understanding of
this resonant effect need to be analyzed the phase factors of the
wavefunctions $\left\vert \varphi _{\pm }(t)\right\rangle $ in Eq. (\ref{TW}%
). These equations show clearly that the wavefunctions $\left\vert \varphi
_{\pm }(t)\right\rangle $ have the nontrivial periodic phase factors $\alpha
_{\pm }(t)=-\Omega _{p}g_{\pm }\sin \omega t/\omega $ apart from the dynamic
phase factors $d_{\pm }(t)=g_{\pm }^{2}t/\omega $ when $\left\vert z_{\pm
}(0)\right\rangle \rightarrow \left\vert z_{\pm }(t)\right\rangle $ driven
by $H_{p}=\Omega _{p}(a^{\dagger }+a)$. When $t=2n\pi $ ($n=1,2,3,\cdots $),
these periodic phase factors $\alpha _{\pm }(t)$ disappear, whereas the
dynamic phase factors $d_{\pm }(t)$ still exist. Importantly, when
calculating the quantum dynamics of $\left\langle \sigma
_{z}(t)\right\rangle =-\left\langle \varphi _{+}(t)\right\vert \sigma
_{z}\left\vert \varphi _{-}(t)\right\rangle -\left\langle \varphi
_{-}(t)\right\vert \sigma _{z}\left\vert \varphi _{+}(t)\right\rangle $,
these periodic phase factors in different subspaces generate a novel
interference. As a result, the term $\xi \sin (\omega t)/2$ in Eq. (\ref{FDT}%
) can be obtained. On the other hand, the nonorthogonal coherent states $%
\left\langle z_{+}(t)\right. \left\vert z_{-}(t)\right\rangle $ and $%
\left\langle z_{-}(t)\right. \left\vert z_{+}(t)\right\rangle $ with the
complex parameters $\xi _{\pm }=-g_{\pm }/\omega +\Omega _{p}\exp (-i\omega
t)$ also leads to another term $\xi \sin (\omega t)/2$ since $D(\xi
_{+})D(\xi _{-})=D(\xi _{+}+\xi _{-})\exp [\frac{1}{2}(\xi _{+}\xi
_{-}^{\ast }-\xi _{-}\xi _{+}^{\ast })]$.

Although the case $\varepsilon =0$ has been realized in the bichromatically
excited of the trap ions \cite{Molmer} and in the spin-orbit-coupled
Bose-Einsten condensate with a harmonic trapped potential \cite{Lin1}, it is
very necessary to discuss the case of $\varepsilon \neq 0$ in the cavity
quantum electrodynamics. If the atomic resonant frequency $\varepsilon $ is
taken into account, Hamiltonian $H_{P}$ is not integrable and the analytical
dynamics, which is similar to Eq. (\ref{FDT}), can not been derived. In Fig.
3, the quantum dynamics of $\left\langle \sigma _{z}(t)\right\rangle $ and
its mean value $M$ for a long time are plotted numerically for $\varepsilon
=0.1\omega $, whereas the other parameters are the same as those in Figs.
(1) and (2). Due to the existence of the term $\varepsilon \sigma _{z}/2$,
the quantum dynamics of $\left\langle \sigma _{z}(t)\right\rangle $ given in
the insert part of Fig. 3 becomes more complicate. However, the quantum
resonance effect with the same positions has still been manifest.

\begin{figure}[tp]
\includegraphics[width=5.0cm]{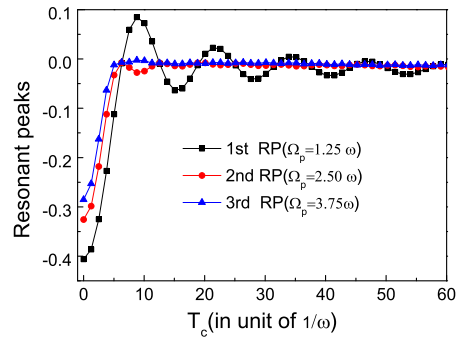}\newline
\caption{(Color online) The first (1st, $\Omega _{p}=1.25\protect\omega $,
black line), second (2nd, $\Omega _{p}=2.50\protect\omega $, red line) and
third (3d, $\Omega _{p}=3.75\protect\omega $, blue line) resonant peaks
(RPs) of the mean value $M$ for a long time $T_{L}=50\protect\pi /\protect%
\omega $ as a function of the rise time $T_{c}$ when $g=0.2\protect\omega $
and $\protect\varepsilon =0$. }
\label{fig4}
\end{figure}
Having obtained the fundamental properties of our predictions, we briefly
address the possible observation in current experimental setup with the
ultrastrong coupling. In principle, our predictions may be observed if the
driving magnitude of the photon field is sufficiently large. As an example,
we here consider the flux-based circuit quantum electrodynamics, in which
the coupling strength $g=2\pi \times 314$ MHz has been reported for $\omega
=2\pi \times 2.782$ GHz \cite{Niemczyk}. This large coupling rate $(g/\omega
\sim 0.11)$ allows us to enter the anticipant ultrastrong coupling regime.
If using ratio $g/\omega =0.1$, the driving magnitudes for the resonant
positions are given by $\Omega _{p}^{m}=2.5\omega $, $5.0\omega $,$\cdots $.
With the increasing of the coupling strength in the near future, the
required driving magnitues become weaker (for example, $\Omega
_{p}^{m}=1.25\omega $, $2.5\omega $,$\cdots $ for $g/\omega =0.2$ in our
calculations and $\Omega _{p}^{m}=0.5\omega $, $1.0\omega $,$\cdots $ for $%
g/\omega =0.5$). On the other hand, for the photon frequency $\omega =2\pi
\times 2.782$ GHz, the integral time $T_{L}$ in Figs. (2) and (3) is given
by $T_{L}=50\pi /\omega = $ $11$ ns, which are shorter than the decoherence
time $t_{D}=1/\kappa =$ $64 $ ns for $\kappa =2\pi \times 2.5$ MHz. It means
that in the range of decoherence time, our predictions can be detected by
measuring the excited state population \cite{MB}. It should be pointed out
that in all simulations above, the driving $H_{p}=\Omega _{p}(a^{\dagger
}+a) $ is turned instantaneously on, which is a useful theoretical
approximation. In fact, the resonant effect is very sensitive to the finite
rise-time of the driving. In Fig. 4, we plot the first, second, and third
resonant peaks of the mean value $M$ for the time $T_{L}=50\pi /\omega $ as
a function of the rise time $T_{c}$, in which the driving magnitude is
chosen as $\Omega (t)=t\Omega _{p}/T_{c}$ for $t<T_{c}$ and $\Omega
(t)=\Omega _{p}$ for $t>T_{c}$. It can be found clearly in the figure that,
if the rise time $T_{c} $ is long, the resonant effect can not be observed.

It is straightforward to find in quantum optics that the spin-boson model
can be also driven by a time-independent field of atom $H_{a}=\Omega
_{a}\sigma _{x}$ with $\Omega _{a}$ being the driving magnitude. In the
strong coupling governed by the Jaynes-Cummings model, these drivings have
almost the same effects on its dynamics. However, in the framework of the
ultrastrong coupling, they generate quite different effects if their driving
magnitudes are strong, as will be demonstated below. Without the atomic
resonant level $(\varepsilon =0)$, the corresponding Hamiltonian is given by
$H_{A}^{\varepsilon =0}=\omega a^{\dagger }a+g\sigma _{x}(a^{\dagger
}+a)+\Omega _{a}\sigma _{x}$. Similar to Eq. (\ref{TW}), the wavefunctions
for Hamiltonian $H_{A}^{\varepsilon =0}$ in the subspaces $\sigma _{x}=\pm 1$
is obtained, if the initial states is chosen as $\left\vert G_{\pm
}(0)\right\rangle $, by $\left\vert \varphi _{\pm }(t)\right\rangle
_{a}=e^{i(\frac{g^{2}}{\omega }t\pm \Omega _{a}t)}\left\vert z_{\pm
}(0)\right\rangle \otimes \left\vert \pm \right\rangle $. In this
wavefunctions, no periodic phase factor can be found. Moreover, the coherent
states remain invariant with respect to $t$. Based on the wavefunctions $%
\left\vert \varphi _{\pm }(t)\right\rangle _{a}$, the quantum dynamics of $%
\left\langle \sigma _{z}(t)\right\rangle _{a}$ is also solved exactly by $%
\left\langle \sigma _{z}(t)\right\rangle _{a}=-\exp (-\frac{4g^{2}}{\omega }%
)\cos (2\Omega _{a}t)$, which is quite different from Eq. (\ref{FDT})
arising from the driving photon field. In this driving, the period of Rabi
oscillation depends only on the driving magnitude $\Omega _{a}$. More
importantly, no resonant effect can be found, even if the driving magnitude $%
\Omega _{a}$ is very strong.

In summery, simulated by the recent experiments of the solid-state quantum
optics with the ultrastrong coupling, we have investigated the simple
quantum dynamics to reveal the fundamental property of the spin-boson model,
in which the counter-rotating terms must be considered. Our predications may
be observed if the driving the photon field is sufficiently strong in
experiments.

We thank Profs. J. -Q. Liang, Chuanwei Zhang, Jing Zhang, and Shiqun
Zhu as well as Drs. Yongping Zhang, Jie Ma and Ming Gong for their
helpful discussions and suggestions. This work was supported partly
by the NNSFC under Grant Nos. 10904092, 10934004, 60978018, 11074184
and 11074154, and the ZJNSF under Grant No. Y6090001.


\begin{thebibliography}{99}
\bibitem{Note} For simplicity, the initial values $c_{+}(0)$ and $c_{-}(0)$
are chosen to be equal. For different initial values, only a phase factor is
generated, and thus, the main results in this Letter are not affected.
\end{thebibliography}
\end{document}